\documentstyle[a4,12pt,epsfig]{article}

\newcommand {\lan}    	{\langle}
\newcommand {\ran}    	{\rangle}
\newcommand {\for}	{\mbox{ for }}
\newcommand {\ie}       {{\em i.e.}\ }
\newcommand {\eg}       {{\em e.g.}\ }
\newcommand {\qea}	{\mbox{$q_{\rm EA}$}}
\renewcommand {\d}    	{\mbox{ d}}

\begin{document}

\title{Generalized off-equilibrium fluctuation-dissipation relations
in random Ising systems}

\author{
Giorgio Parisi\thanks{e-mail: {\tt
giorgio.parisi@roma1.infn.it}},\ \ 
Federico Ricci-Tersenghi\thanks{e-mail: {\tt
federico.ricci@roma1.infn.it}} \\
{\small $^1$ Dipartimento di Fisica and INFN, Universit\`a di Roma I
{\it La Sapienza}}\\
{\small P. A. Moro 2, 00185 Roma, Italy}\\
\and
Juan J. Ruiz-Lorenzo\thanks{e-mail: {\tt
ruiz@lattice.fis.ucm.es}}\\
{\small $^2$ Departamento de F\'{\i}sica Te\'orica I, Universidad
Complutense de Madrid}\\
{\small 28040 Madrid, Spain}
}

\date{26 November 1998}

\maketitle

\abstract{
We show that the numerical method based on the off-equilibrium
fluctuation-dissipation relation does work and is very useful and
powerful in the study of disordered systems which show a very slow
dynamics.  We have verified that it gives the right information in the
known cases (diluted ferromagnets and random field Ising model far
from the critical point) and we used it to obtain more convincing
results on the frozen phase of finite-dimensional spin glasses.
Moreover we used it to study the Griffiths phase of the diluted and
the random field Ising models.
}

\baselineskip = 21pt

\section{Introduction}
\label{S_INT}

In the last years more and more interest has been devoted to systems
which show a very slow dynamics~\cite{OUT}. Major examples are spin
glasses and structural glasses, which can be viewed as a supercooled
liquid which slowly evolves never reaching equilibrium.  In the study
of such systems is clear the important role played by the
off-equilibrium dynamics, which describes the behavior of the real
system, in contrast with the equilibrium thermodynamic properties
(which may differ from that of a not well thermalized
system)~\cite{BOOK}.

Two well known spin systems which show very slow approach to
equilibrium are spin glasses~\cite{YOUNG_BOOK,MEPAVI} and diluted
ferromagnets. In both cases it has been found that the effective
dynamical exponent $z$ in the frozen phase is quite large. Such effect
makes the dynamics of those models very slow in the frozen phase.

In the last years we have used and compared two different kinds of
simulations: a) Monte Carlo simulations by which we measure the {\em
statics} of the model, \ie we sample the configurational space
according to the usual equilibrium Gibbs-Boltzmann distribution; b) a
second kind of Monte Carlo during which we keep the system in the out
of equilibrium regime, we let it relax during a very large waiting
time and then we measure quantities slowly varying in time (the {\em
dynamics} of the model). Hereafter with the word 'dynamical' we will
always refer to the out of equilibrium dynamics and never to the
dynamics at the equilibrium.  Thanks to some relations which link the
thermodynamical observables to these off-equilibrium quantities we are
able to calculate, for the Edwards-Anderson (EA) model, the critical
exponents measuring the relaxational dynamics at the critical
point~\cite{6DIM}, and, as we will show in the present work, also the
(functional) order parameter in the frozen phase.

The advantages of these dynamical studies are manifold.  We do not
need to thermalize the system (which is a very hard task in disordered
systems) and so we are not restricted to study very small volumes. On
the contrary we must use very large volumes to keep the sample in the
off-equilibrium regime [$\xi(t) \ll L$, where $\xi(t)$ is the
dynamical correlation length, which is the typical distance over which
the system is equilibrated at time $t$] and therefore the finite size
effects are irrelevant for not too large times.  We can obtain
information on the statics of the model via the dynamical scaling and
we can predict the value of not self-averaging quantities (like the
$P(q)$ in spin-glasses) only measuring self-averaging ones, and so we
do not need to average over a large number of disorder
realizations. This very important feature could also be exploited by
the experimentalists to calculate the distribution function of the
overlap in a spin glass sample, which until today was measurable only
in numerical simulations~\footnote{An indirect experimental
determination of $P(q)$ has been done in~\cite{ORBACH}.}.

In this work we present the results of a dynamical study on the
diluted ferromagnetic and the random field models on one side and the
spin glass model on the other.  These models behave very differently
in their statics, even if they age similarly in the out of equilibrium
regime.  Using the numerical method based on the off-equilibrium
fluctuation-dissipation relation (OFDR), introduced for the study of
the EA model by Franz and Rieger~\cite{FRARIE} and summarized in the
next section, we can measure the correct equilibrium properties of the
model.  The presence of a large number of metastable states in all the
models does not invalidate the method, which succeed in predicting the
right thermodynamical state.  Once checked the validity of the method,
we can use it to determine whether the frozen phase of the spin glass
model in finite dimensions is better described by the mean-field like
solution\cite{MEPAVI,BOOK,PRL3D,6DIM,4DIM,FDT} or by the droplet
model~\cite{DROPLET}.

Few analytical results and the definitions of the models studied will
be presented in the next section. In the third and fourth sections we
show the numerical results and in the last one our conclusions.  In
the last year the OFDR has been measured in many different systems.  A
clear classification of the models seems to come out from these
studies.  This classification will be presented in the Appendix.

\section{Analytical results}
\label{S_ANA}

\subsection{Off-equilibrium fluctuation-dissipation relation}

Assuming time translation invariance (TTI), which is valid in the
evolution of a system at the equilibrium, can be proved the
fluctuation-dissipation theorem (FDT), that reads
\begin{equation}
R(t,t^\prime)=\beta\ \theta(t-t^\prime)\ \frac{\partial
C(t,t^\prime)}{\partial t^\prime} \ ,
\label{FDT}
\end{equation}
where $\beta$ is the inverse temperature and $\theta(t-t^\prime)$ is
the step function, given by causality.  Here TTI implies that
$C(t,t^\prime) = {\cal C}(t-t^\prime)$ and $R(t,t^\prime) = {\cal
R}(t-t^\prime)$.  The autocorrelation (a two times function) and the
response function are defined as follows
\begin{equation}
C(t,t^\prime) \equiv \overline{\lan A(t) A(t^\prime) \ran} \quad ,
\quad R(t,t^\prime) \equiv \left. \frac{\overline{\lan \delta
A(t)\ran}}{\delta \epsilon(t^\prime)}\right|_{\epsilon=0} \ ,
\end{equation}
where we assume that the original Hamiltonian has been perturbed by a
term
\begin{equation}
{\cal H}^\prime= {\cal H} + \int \epsilon(t) A(t) \ .
\end{equation}
A common choice in spin models is $A(t) = \sum_i \sigma_i(t)$ and
$\epsilon(t)$ as an external magnetic field.  The brackets $\lan
(\cdot\cdot\cdot) \ran $ imply here an average over the dynamical
process and the overbar $\overline{(\cdot\cdot\cdot)}$ a second one
over the disorder.

In the out of equilibrium evolution (starting for example from a
random configuration) the FDT does not hold any more.  Nonetheless
some years ago has been proposed by Cugliandolo and
Kurchan~\cite{CUKU} a generalization of this theorem which should be
valid in the early times of the dynamics, when the system is too far
from the equilibrium.  Such generalization, that we call
off-equilibrium fluctuation-dissipation relation (OFDR), has been
obtained in the study of spin glass mean-field models.  Only recently
has been numerically verified that such generalization is also valid
in a short-range spin glass model~\cite{FDT}.

Let us recall how this relation can be obtain.  In the off-equilibrium
regime TTI is no longer valid and so all two-times functions depend
explicitly on both times and not only on their difference.  The
fundamental assumption~\cite{CUKU} is that, in this regime, FDT is
modified simply by a multiplicative factor $X(t,t^\prime)$.  This
assumption have been verified in Ref.~\cite{FDT} for the EA model.  In
the large times region ($t \gg 1, t^\prime \gg 1$) the violation
factor depends on the times $t$ and $t^\prime$ only via the
autocorrelation function~\cite{BCKP}:
$X(t,t^\prime)=X[C(t,t^\prime)]$.

Assuming OFDR
\begin{equation}
R(t,t^\prime)=\beta\ X[C(t,t^\prime)]\ \theta(t-t^\prime)\
\frac{\partial C(t,t^\prime)}{\partial t^\prime}\ ,
\label{OFDR}
\end{equation}
we can now extract the violation factor $X(C)$ from the measures of
autocorrelation and magnetization, which are self-averaging
quantities.

In the linear-response regime ($h \ll 1$) we can write the
magnetization as
\begin{equation}
m[h](t) = \int_{-\infty}^t \d t^\prime\ R(t,t^\prime) h(t^\prime) \ ,
\label{lin-res}
\end{equation}
where the upper limit of the integral has been set to $t$ due to
causality.

Substituting Eq.(\ref{OFDR}) into Eq.(\ref{lin-res}) we have that
\begin{equation}
m[h](t) = \beta \int_{-\infty}^t \d t^\prime\ X[C(t,t^\prime)]
\frac{\partial C(t,t^\prime)}{\partial t^\prime} h(t^\prime) \ .
\end{equation}
The way we perturb the system is important for the numerical
purposes\footnote{In the first work on the subject Franz and
Rieger~\cite{FRARIE} used $h_i(t)=h\ \theta(t_w-t)$, that gives less
clear results.} and we choose to switch on a random field of intensity
$h_i$, that depends on the site, at time $t_w$: $h_i(t)=h_i\
\theta(t-t_w)$, where $h_i$ is a Gaussian variable with zero mean and
variance $h_0$.  The corresponding magnetization and susceptibility
are defined by
\begin{eqnarray}
m[h](t) = \overline{\lan \sigma_i h_i \ran}/h_0 \quad ,\\
\chi(t,t_w) = \lim_{h_0 \to 0} \frac{m[h](t)}{h_0} \quad .
\end{eqnarray}
Then we have that
\begin{equation}
\chi(t,t_w) = \beta \int_{t_w}^t \d t^\prime\ X[C(t,t^\prime)]
\frac{\partial C(t,t^\prime)}{\partial t^\prime}\ ,
\end{equation}
and by performing the change of variables $u=C(t,t^\prime)$ we finally
obtain the key equation
\begin{equation}
\chi(t,t_w) = \beta \int_{C(t,t_w)}^{1} \d u\ X(u) \ ,
\label{key}
\end{equation}
where we have used the fact that $C(t,t) \equiv 1$ in Ising models.

From Eq.(\ref{key}) the violation factor $X(C)$ can be easily
extracted simply measuring the autocorrelation function $C(t,t_w)$ and
the integrated response to a small external field $\chi(t,t_w)$.

Eq.(\ref{key}) can be rewritten as
\begin{equation}
T \chi(t,t_w) = S[C(t,t_w)] \ ,
\end{equation}
where we have defined
\begin{equation}
S(C) = \int_{C}^{1} \d u\ X(u) \ .
\end{equation}
We remark that all the information we need is encoded in the shape of
the function $S(C)$.

If the system is at equilibrium the violation factor is equal to one
and the relation becomes
\begin{equation}
T \chi(t,t_w) = 1 - C(t,t_w) \qquad \mbox{or} \qquad
S(C) = 1-C \ .
\end{equation}

\subsection{Link between the statics and the dynamics}

To get information on the thermodynamical properties of the model we
should match the violation factor $X(C)$ to some static observable.
This can be done using the following conjecture~\cite{CUKU,FM} (proved
under some assumptions in~\cite{FRMEPAPE}) on the large times behavior
of the $X(C)$, which has already been verified in~\cite{FDT}.  Sending
$t \to \infty$ and $t_w \to \infty$, keeping $C(t,t_w)=q$
\begin{equation}
X[C(t,t_w)] \longrightarrow x(q) \ .
\end{equation}

The function $x(q)$ is well known in spin glass theory~\cite{MEPAVI}
and it is linked to the thermodynamical order parameter $P(q)$ via
\begin{equation}
x(q)=\int_{0}^q \d q^\prime\ P(q^\prime)\ ,
\end{equation}
where the overlap distribution function $P(q)$ is the thermodynamic
probability of finding two copies of the system with overlap $q$.
Moreover, we can define
\begin{equation}
s(C) = \int_{C}^{1} \d u\ x(u) \ .
\end{equation}

Now we have all the ingredients for our numerical recipe: we measure
the autocorrelation function and the integrated response to a small
external field for large times, then we make a derivation and we
obtain the function $P(q)$.  The meaning of the outcoming function
$P(q)$ is well described in~\cite{MEPAVI}.

A classification of the models can be given in terms of the violation
factor $X$.  We will report this classification in the Appendix.

\section{The models}
\label{S_MODELS}

We simulate Ising spin models on cubic ($d=3$) or hypercubic ($d=4$)
lattices with the following Hamiltonian
\begin{equation}
{\cal H} = -\sum_{<ij>} J_{ij} \sigma_i \sigma_j -\sum_i h_i
\sigma_i \quad ,
\end{equation}
where the first sum runs over the first-neighbors pairs. Depending on
the model the couplings $J_{ij}$ and the fields $h_i$ are fixed to
some value or are taken randomly with some distribution function.

The three random models we are interested in are:
\begin{description}
\item[SDIM] (Site-Diluted Ising Model) where $J_{ij} = J \epsilon_i
\epsilon_j$ and $h_i = 0$. Every $\epsilon_i$ follows the probability
distribution $P(\epsilon_i) = c\:\delta(\epsilon_i-1) +
(1-c)\delta(\epsilon_i)$, where $c$ is the spin concentration.  The
dimensionality will be $d=3$.
\item[RFIM] (Random Field Ising Model) where $J_{ij} = 1$ and every
$h_i$ follows the probability distribution
$P(h_i)=\frac{1}{2}[\delta(h_i+h_r)+ \delta(h_i-h_r)]$.  The
dimensionality will be $d=3$.
\item[EA] (Edwards-Anderson Model) where $J_{ij}$ is a Gaussian
distributed variable of zero mean and unit variance.  Moreover, in
this case, we have set the magnetic field to zero ($h_i=0$). We will
study this model in four dimensions.
\end{description}

\section{Numerical results}
\label{S_NUM}

All the simulations have been performed on a {\em tower} of the
parallel computer APE100~\cite{APE}, with a peak performance of 25
Gigaflops.

We have numerically computed the function $S(C)$ for the models
presented in Sect.~\ref{S_MODELS}.  All the plots present the
integrated response multiplied by the temperature versus the
autocorrelation.  These plots should be read from right to left and
from bottom to top, since during the simulation $C(t,t_w)$ starts from
$1$ and falls off, while $\chi(t,t_w)$ starts from zero and increases.
The equation of the line is always $1-C$ and we define $\qea(t_w)$ as
the value of the autocorrelation when the data with waiting time $t_w$
left the straight line $1-C$.  In the model which present a
ferromagnetic transition we often use $m_0^2$ instead of \qea, being
$m_0$ the spontaneous magnetization.

\begin{figure}
\begin{center}
\leavevmode
\epsfxsize=0.7 \textwidth
\epsffile{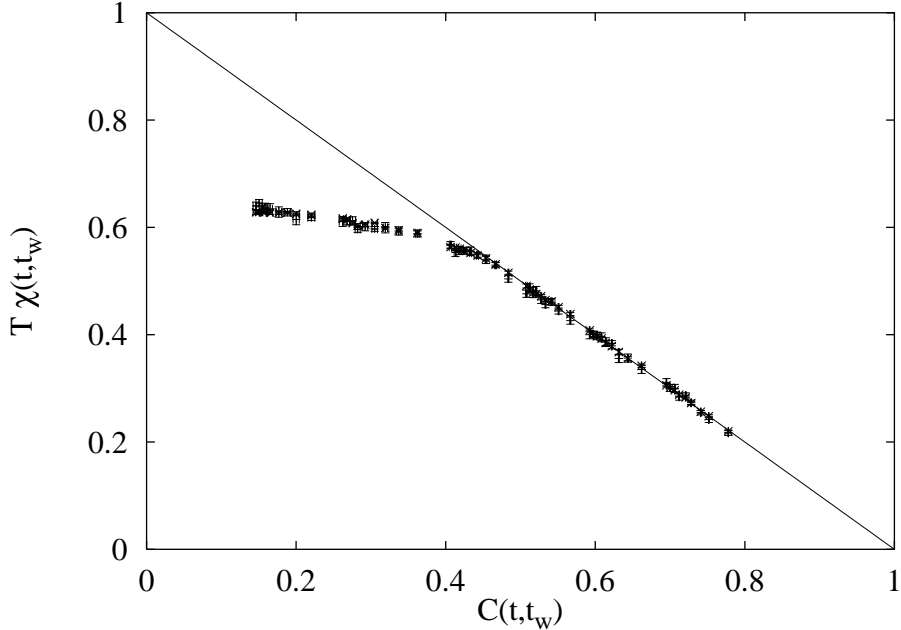}
\end{center}
\caption{This plot shows that we are working in the linear response
regime. The data obtained from the simulation of a spin glass model
($T=1.35=\frac34 T_c$ and $t_w=10^4$) using two different magnetic
fields ($h_0=0.1$ and $h_0=0.05$) used to compute the response
function, are perfectly superimposed.  In the following we will use a
perturbing field of intensity $h_0=0.1$.}
\label{lin_res}
\end{figure}

We have checked that we work in the linear response regime by
simulating all three models with perturbations of intensity $h_0=0.1$
and $h_0=0.05$. We have checked that both magnetic field gives the
same response function.  For example, in Fig.~\ref{lin_res}, the data
for $h_0=0.1$ and $h_0=0.05$ coincide in the four dimensional spin
glass.  Moreover, we always simulated very large volumes with a small
number of disorder realization (typically from 2 to 6), since we are
measuring correlation and response functions, which are
self-averaging.

\subsection{Site-diluted Ising model}

\begin{figure}
\begin{center}
\leavevmode
\epsfxsize=0.7 \textwidth
\epsffile{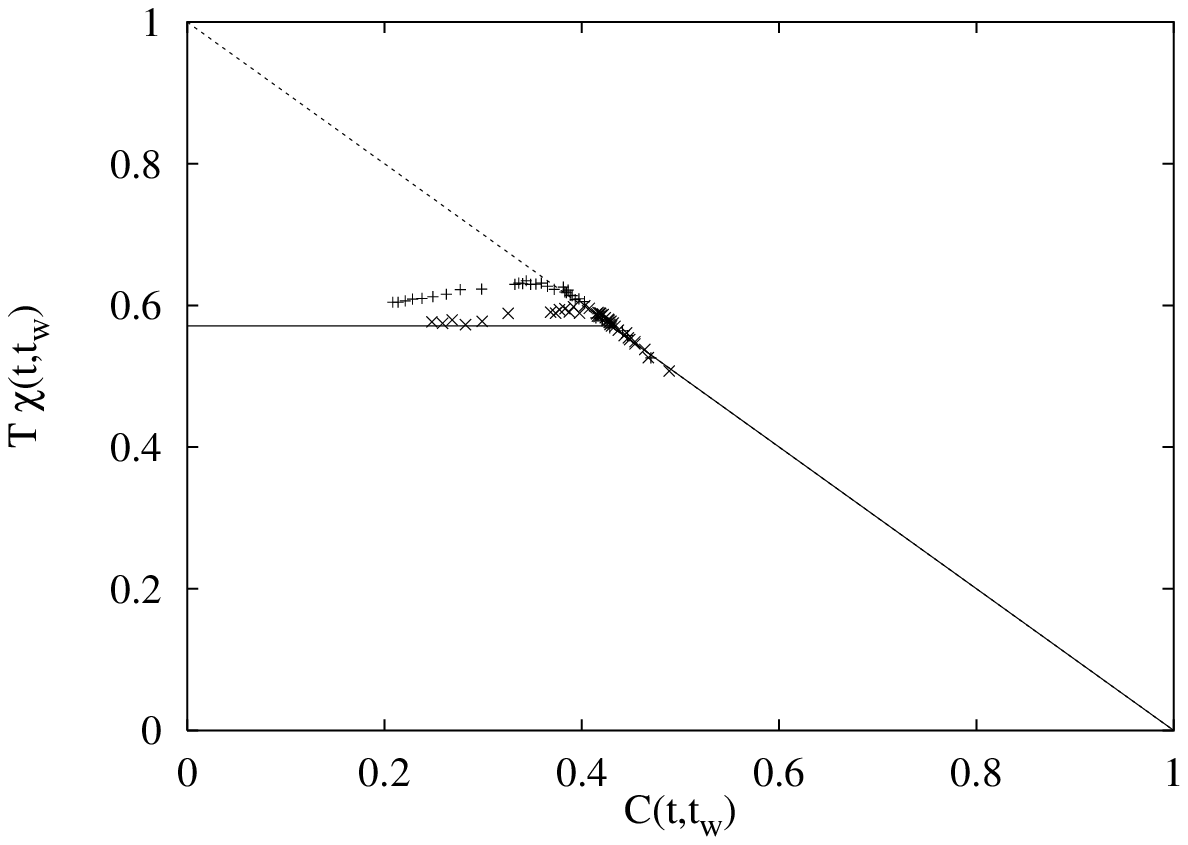}
\end{center}
\caption{The function $S(C)$ for the $3d$ diluted Ising model with a
spin concentration $c=0.65$. We used a volume of $400^3$ and a
temperature $T=2.4314=0.9\ T_c$. The two sets of data have been
measured after a waiting time $t_w=10^3$ (uppermost) and $t_w=10^4$
(lowermost). The line $1-C$ is the FDT regime, while the horizontal
line is the infinite time limit of the data. See the text for more
details.}
\label{dilu065}
\end{figure}

\begin{figure}
\begin{center}
\leavevmode
\epsfxsize=0.7 \textwidth
\epsffile{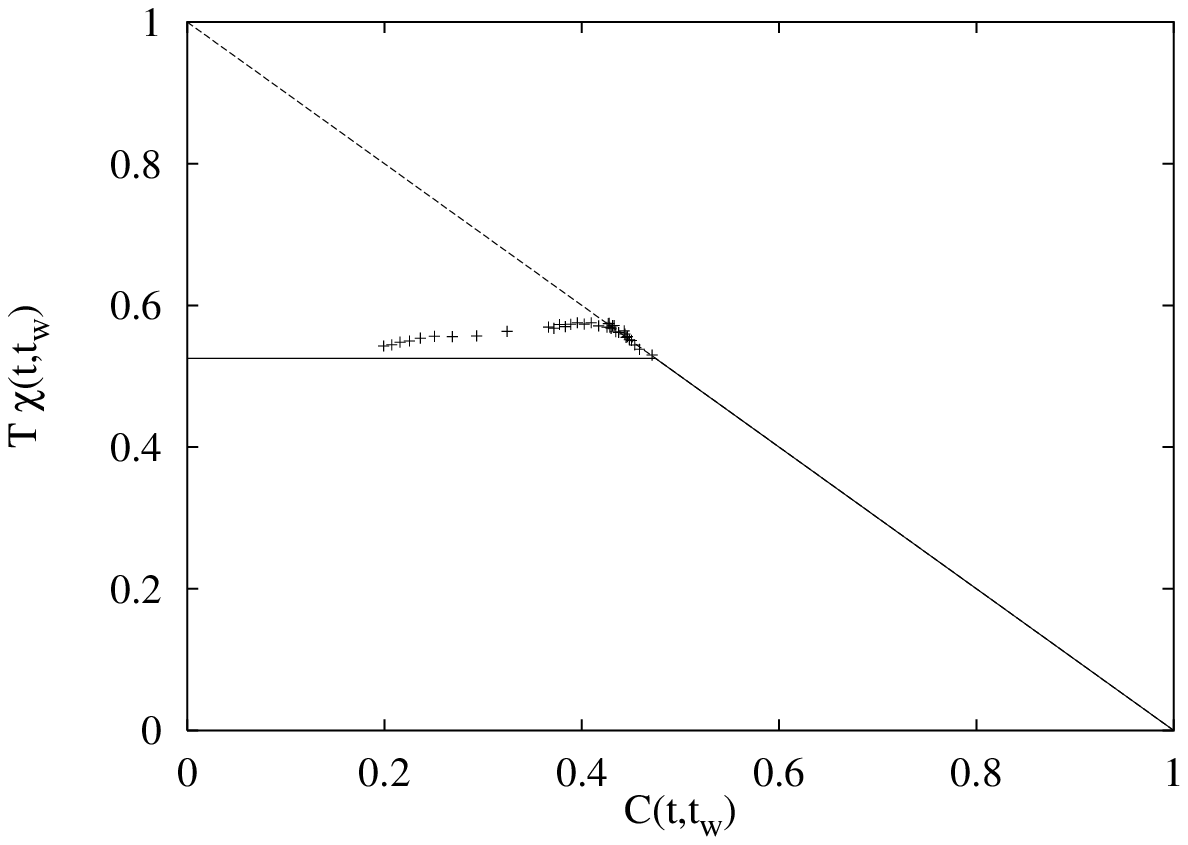}
\end{center}
\caption{The function $S(C)$ for the $3d$ diluted Ising model with a
spin concentration $c=0.8$. The volume is $520^3$, the temperature
$T=3.1497=0.9\ T_c$ and $t_w=10^3$. The line $1-C$ is the FDT regime,
while the horizontal line is the infinite time limit of the data. See
the text for more details.}
\label{dilu08}
\end{figure}

In Fig.~\ref{dilu065} and \ref{dilu08} we show the data for the
site-diluted ferromagnetic model in $d=3$, with concentration
respectively $c=0.65$ and $c=0.8$.

A great care has to be taken in order to keep the system in the out of
equilibrium situation in the whole run. We require the off-equilibrium
correlation length $\xi(t)$ ~\cite{PRL3D,4DIM,6DIM} to be always much
less than the sample size $L$. This requirement is quite easy to
satisfy in a spin glass due to the high value of the dynamical
exponent $z$: working in the frozen phase ($T \simeq 3/4\;T_c$) is
practically impossible to thermalize a sample with $L=32$, which is
the size that we use in the simulations described in
Sect.~\ref{SS_EA}.  In a diluted ferromagnet the situation is much
more subtle because two main reasons. Firstly the dynamics is slow but
not so much like in spin glasses, especially for small dilution (on
the time scales we use the ``effective'' dynamical exponent $z$ seems
to increase with the dilution~\cite{HEUER,Z_EFF}).  On the other hand,
we must simulate in the frozen phase, where the spontaneous
magnetization $m_0$ is different from zero, but near to the critical
temperature in order to have a small order parameter.

To solve this problem we have used always large volumes (up to 140
millions of spins) and we have checked how far from equilibrium the
system was during all the simulation.  The simplest way to do that is
to measure the absolute value of the instantaneous magnetization,
which is very small in the off-equilibrium dynamics and which grows,
around the thermalization time, converging to $m_0$. All the data
presented here come out from runs where the magnetization is
statistically compatible with zero in the whole simulation.

Two different regimes can be clearly distinguished.  In the regime
$(t-t_w)<t_w$, or equivalently $C(t,t_w)>m_0^2(t_w)$ the data stay on
the line $T \chi(t,t_w) = 1-C(t,t_w)$, which means that the system is
in a quasi-equilibrium regime where the relation between the
correlation and the response is like that at the equilibrium.  In the
regime $(t-t_w)>t_w$ [\ie $C(t,t_w)<m_0^2(t_w)$] the data leave the
straight line and the violation of the FDT can be summarized in the
$X$ factor, which is nothing but the derivative of the curve followed
by the data (with the opposite sign).

It is clear from the plots that the diluted ferromagnetic model
belongs to the category A defined in the Appendix, with $X=0$ and the
data which stay on a horizontal line.

In Fig.~\ref{dilu065} and~\ref{dilu08} we draw also the lines
corresponding to the infinite time limit.  As explained in the
Appendix, in the ferromagnetic phase the model should have an
\begin{equation}
s(C) = \left\{
\begin{array}{lll}
1-C & \quad \for & C>m_0^2 \quad ,\\
1-m_0^2 & \quad \for & C\le m_0^2  \quad ,\\
\end{array}
\right .
\end{equation}
being $m_0$ the equilibrium spontaneous magnetization.

The spontaneous magnetization have been calculated from equilibrium
simulations, in order to check the convergence of the data measured in
the out of equilibrium regime.  As it can be seen the data presented
are already very near to the asymptotic regime.  The horizontal line
in Figures~\ref{dilu065} and~\ref{dilu08} is given by $1-m_0^2$.  For
our purposes the main information is that the shape of $S(C)$ does not
depend strongly on the value of $t_w$ in a large time range.

A further remark on these data is needed in order to justify why
$\chi(t,t_w)$ slightly decreases with time in the region
$(t-t_w)>t_w$. Being $\chi(t,t_w)$ the integral of the response, which
we would expect that is a non-negative function, it is surprising that
it is a decreasing function of the time.  The little decrease is
mainly due to the high susceptibility of the spins placed on the
interfaces (domain walls), which give to the response an extra
contribution which will disappear for longer times, when the fraction
of spins on the domain boundaries gets smaller.  This effect has been
already found by A.~Barrat~\cite{BARRAT} in the study of the OFDR in a
coarsening pure ferromagnet.

The results from the diluted ferromagnetic model are very important
because they rule out the possibility that what we measured in spin
glasses~\cite{FDT} was simply a dynamical artifact, which shadows the
true static behavior.  Now we are sure that the method based on the
OFDR is able to give us information on the right thermodynamic state,
no matter of how complex is the dynamics followed by the system in
reaching that state.

\begin{figure}
\begin{center}
\leavevmode
\epsfxsize=0.7 \textwidth
\epsffile{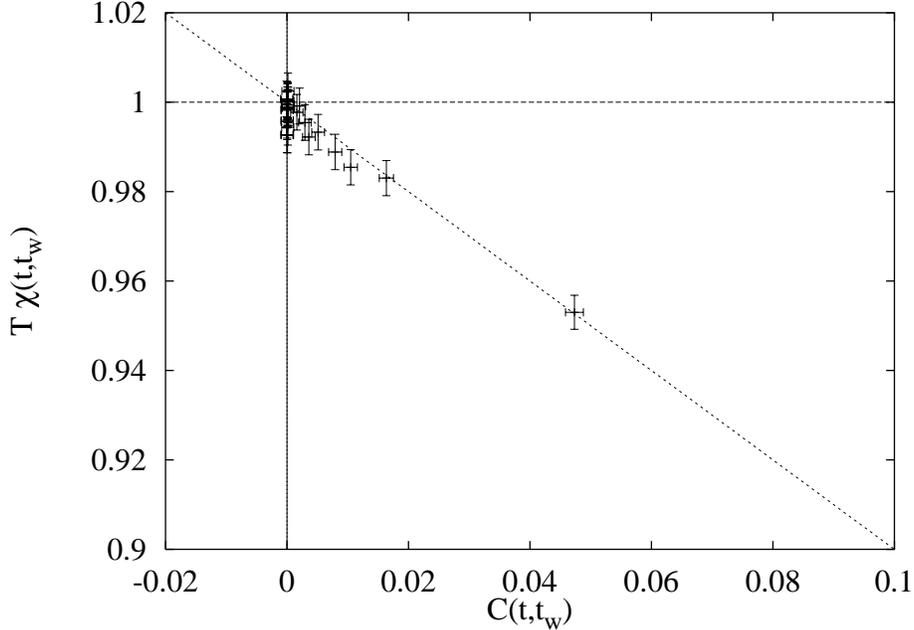}
\end{center}
\caption{The $3d$ diluted Ising model with $c=0.8$ and $T=4.0$, that
is in the region, $T_c \simeq 3.5 < T < T_c^{(0)} \simeq 4.5$, where
the Griffiths singularities~\protect\cite{GRIFFITHS} may break the
replica symmetry, clearly behaves like in the paramagnetic and replica
symmetric phase (\ie the data always stay on the FDT line) with the
autocorrelation decreasing to zero in a fast way.  The results do not
depend on $t_w$.  Note that the plot range is zoomed in the upper left
corner of the usual range.}
\label{dilu_high}
\end{figure}

One more result from these simulations is the strong hint that the
diluted ferromagnetic model has no replica symmetry breaking in the
Griffiths phase~\cite{GRIFFITHS}. 
To this purpose a run was also performed in the
temperature region between the critical temperature of the pure model
($T_c^{(0)} \simeq 4.5$~\cite{DILU3D}) and that of the diluted one
($T_c \simeq 3.5$~\cite{DILU3D}).  In this region the model could show
a replica symmetry breaking~\cite{DOTSENKO}.  Our results
(Fig.~\ref{dilu_high}) show very clearly that there is no replica
symmetry breaking and the data behaves in the same way they do in the
paramagnetic region, \ie they stay on the FDT line, while the
autocorrelation decays rapidly to the limiting value, which is zero
for the diluted ferromagnetic model in the high temperature phase.

A heuristic analytical argument can be provided in order to justify
why in a diluted ferromagnetic model (with $Z_2$ as global symmetry) 
can not exist a phase transition
from a paramagnetic to a spin glass phase.  In the paramagnetic phase
of a ferromagnetic model (diluted or not) the two-point correlation
function is positive and
obviously less than 1
\begin{equation}
0 < \lan \sigma_i \sigma_j \ran \le 1 \qquad \forall\ i,j \ .
\end{equation}
Using the definitions of the magnetic and spin glass susceptibility
for $T>T_c$
\begin{equation}
\chi_m = \sum_{i,j} \overline{ \lan \sigma_i \sigma_j \ran }\qquad
\chi_q = \sum_{i,j} \overline{ \lan \sigma_i \sigma_j \ran^2} \ ,
\end{equation}
where the angular brackets $\langle (\cdot\cdot \cdot)\rangle$ stands
for the thermal equilibrium average in a given sample and
$\overline{(\cdot \cdot \cdot)}$ stands for average over the disorder.
The inequality $\lan \sigma_i \sigma_j \ran^2 \le \lan \sigma_i
\sigma_j \ran$ can be easily demonstrated, and so, above the critical
temperature we have that
\begin{equation}
\chi_q \le \chi_m \qquad \forall\ T>T_c \ .
\label{E-chi}
\end{equation}
The transition to a spin glass phase is usually defined as the
divergence of the spin glass susceptibility ($\chi_q$), while the
magnetic one ($\chi_m$) remains finite.  This can not happen if
Eq.(\ref{E-chi}) holds.

\subsection{Random field Ising model}

\begin{figure}
\begin{center}
\leavevmode
\epsfxsize=0.7 \textwidth
\epsffile{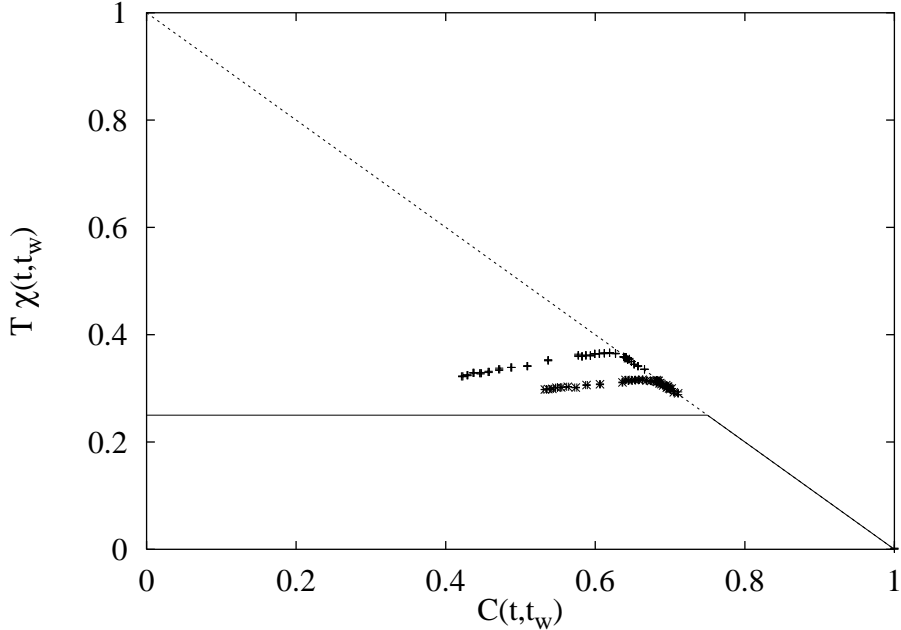}
\end{center}
\caption{The function $S(C)$ for the 3d RFIM with bimodal field
distribution.  The volume is $440^3$, the field $h_r=1.2425$, the
temperature $T=3.2=0.9\ T_c$ and the waiting time $t_w=10^3$.  The
data behave like in a ferromagnet and they should converge to the
horizontal line.}
\label{rfim_low}
\end{figure}

\begin{figure}
\begin{center}
\leavevmode
\epsfxsize=0.7 \textwidth
\epsffile{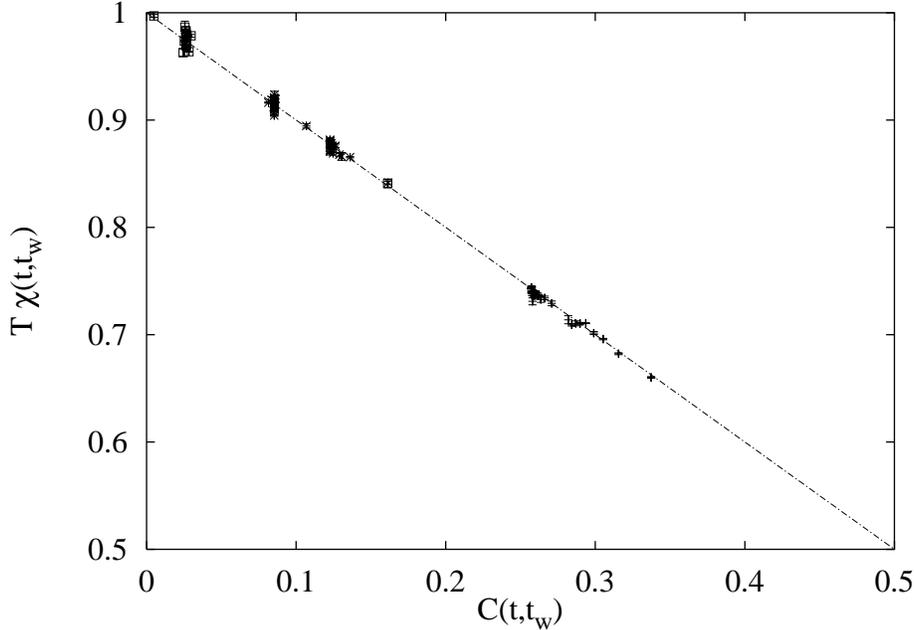}
\end{center}
\caption{High temperature behavior of the RFIM with bimodal field
distribution.  We have used a volume of $440^3$ and a field $h_r=\pm
1.2425$.  The four data spots correspond to temperatures (from below
to top) $T=3.9, 4.5, 5.0, 8.0$.  The results are $t_w$-independent.}
\label{rfim_high}
\end{figure}

In Fig.~\ref{rfim_low} and~\ref{rfim_high} we show the results for the
$3d$ random field Ising model (RFIM) with bimodal field distribution
$h_i=\pm h_r=\pm 1.2425$, whose critical temperature is
$T_c=3.55$~\cite{RFIM_TC}.  We have simulated very large volumes,
$440^3$, in a wide range of temperatures.

The main result in the low temperature phase, $T<T_c$, is that the
function $S(C)$ converges quite rapidly to the right ferromagnetic
equilibrium function (category A of the Appendix):
\begin{equation}
s(x) = \left\{
\begin{array}{lll}
1-x & \quad \for & x>m_0^2 \quad ,\\
1-m_0^2 & \quad \for & x \le m_0^2 \quad ,
\end{array}
\right .
\end{equation}
where $m_0$ is the equilibrium spontaneous magnetization. This
limiting function is plotted with a continuous line in
Fig.~\ref{rfim_low}. The data plotted in that figure are relative to
$t_w=10^2$ (uppermost) and $t_w=10^3$ (lowermost).  The decrease of
the data for $C(t,t_w) < m_0^2(t_w)$ is due to the same reason we
explained before in the case of the diluted Ising model and, as it can
be seen in Fig.~\ref{rfim_low}, the effect tends to disappear for
larger $t_w$.

The data for $T>T_c$ are plotted in Fig.~\ref{rfim_high} and the four
data spots correspond to temperatures, from bottom to top, $T=3.9,
4.5, 5.0, 8.0$.  Note that the abscissa range used is smaller than the
typical one, because the data converges rapidly to the equilibrium
value.  The aging effects quickly disappear and the results become
independent of the value of $t_w$ and so we are able to used always a
quite small value for the waiting time $t_w=10^2$.  Note that in the
RFIM the autocorrelation function does not tend to zero even in the
high temperature range.  This can be simply understood writing a
mean-field equation, that it should work in the high temperature
region even at a quantitative level, for the equilibrium magnetization
of a given sample
\begin{equation}
m_i = \tanh\left(\beta \sum_j m_j + \beta h_i\right) \quad ,
\label{E-mag}
\end{equation}
where $m_i = \langle \sigma_i \rangle$ and the angular brackets
$\langle (\cdot\cdot \cdot)\rangle$ stands for the thermal equilibrium
average in a given sample.  For high temperatures the first term in
the hyperbolic tangent argument in Eq.(\ref{E-mag}) will be smaller
than the second one.  In fact assuming that $m_i \simeq \tanh(\beta
h_i) \propto \beta$ for small $\beta$ values, we have that $\beta
\sum_j m_j \propto \beta^2$, which is much smaller than $\beta h_i$.

For large times the equilibrium autocorrelation function tends to
$\lim_{\tau \to \infty} C(\tau) = N^{-1} \sum_i m_i^2$ (we call this
limit $C_\infty$) and then we have that in the RFIM this limit is
non-zero even in the high temperature region and equals
\begin{equation}
C_\infty(\beta,h_r) = \int \d h P(h) \tanh^2(\beta h) = \tanh^2(\beta
h_r) \quad ,
\label{E-cinf}
\end{equation}
where the last equality hold only for the bimodal random field
distribution we have used.  We have verified that the autocorrelation
function at temperatures $T=5.0$ and $T=8.0$ converges to the
$C_\infty$ value given by Eq.(\ref{E-cinf}), as it should.

Our conclusion are that all the data behave in the same way of the
diluted Ising model, showing no replica symmetry breaking, even in the
temperature region between the critical temperatures of the random
model and that of the pure one (this region defines the Griffiths
phase of the model).  A spin glass phase could be expected just above
the critical temperature~\cite{RFIM_RSB} (which is $T_c \simeq 3.55$
in our case), but from our data measured at $T=3.9 \simeq 1.1\; T_c$
we can conclude that, if this phase exists, it should be in a very
narrow temperature range.  In fact in the work of
Sacconi~\cite{SACCONI} in $d=4$ the spin glass critical temperature
was found to be only few percent greater than the ferromagnetic one.
The region very close to the critical one (\ie $3.55 < T < 3.9$) is
not yet explored and we are actually investigating it.

\subsection{Edwards-Anderson model}
\label{SS_EA}

\begin{figure}
\begin{center}
\leavevmode
\epsfxsize=0.7 \textwidth
\epsffile{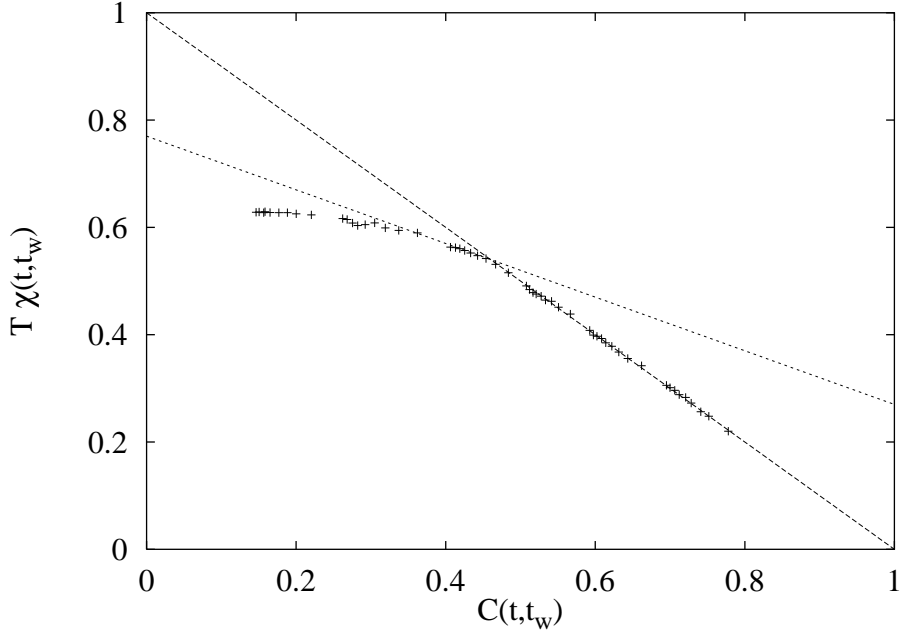}
\end{center}
\caption{The function $S(C)$ for the $4d$ Ising spin glass.  The
volume is $32^4$, the temperature $T=1.35=0.75\ T_c$ and $t_w=10^4$. The
line $1-C$ is the FDT regime, while the other one is only a guide for
the eyes (see text).}
\label{sg}
\end{figure}

The data plotted in Fig.~\ref{sg} are obtained from a simulation of an
Edwards-Anderson (EA) model in $d=4$ spatial dimensions. They are of
the same type of the ones already presented in~\cite{FDT}. In addition
to previous data~\cite{FDT} we have simulated very larges waiting
times. Moreover, we report them to make a comparison between the
models belonging to categories A and C.  Anyhow note that the data
reported here come out from longer runs and now we can assert with
higher confidence that the EA model belongs to category C,
contradicting the droplet theory~\cite{DROPLET} which assign it to
category A.

We tried to fit the data to a two-lines behavior (categories A and B),
but the result was discarded because of the high $\chi^2$ value. The
only prediction which seems to be compatible with our data is the one
coming from the mean-field-like scenario of finite-dimensional spin
glasses~\cite{PRL3D,MAPARU,REVIEW}, which predicts a full replica
symmetry breaking solution for the Edwards-Anderson model and assign
it to category C.  In Fig.~\ref{sg} we also plot a second straight
line to emphasize the curvature of the data in the aging regime.

The data plotted in Fig.~\ref{sg} have $t_w = 10^4$ and we believe
that they have an $S(C)$ function very similar to the asymptotic one.

The time dependence of the fluctuation-dissipation ratio in spin
glasses is a subject to be dealt with very much care, because of the
very slow dynamics.  We dedicated to the study of this subject a great
numerical effort (several weeks of the parallel computer APE100) to be
able to extrapolate our results to the infinite times limit where we
can link it with the statics.  We simulated 12 systems of size $32^4$
with waiting times as large as $t_w=10^6$ and magnetic fields of
intensity $h_0=0.1$ and $h_0=0.05$.  With such very high values of
$t_w$ we can safely extrapolate to infinite waiting times our data.

\begin{figure}
\begin{center}
\leavevmode
\epsfxsize=0.7 \textwidth
\epsffile{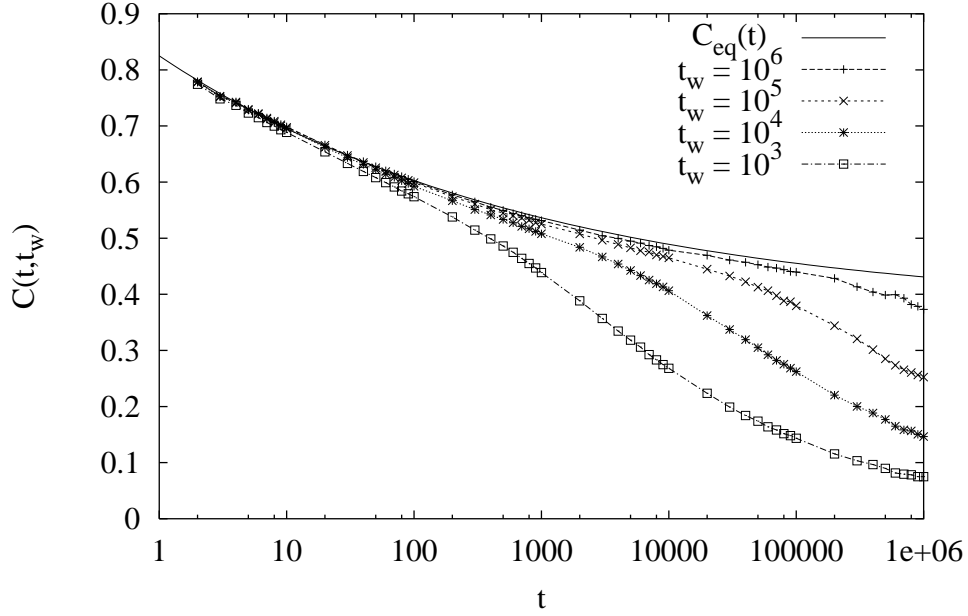}
\end{center}
\caption{The autocorrelation functions of a system of volume $32^4$
and temperature $T=1.35$ with many different waiting times. The higher
curve is the infinite time extrapolation: $C_{\rm eq}(t) = 0.37(1) +
0.455(5)\:t^{-0.145(3)}$.}
\label{corr}
\end{figure}

To confute the prediction of the droplet theory that assign the EA
model to category A, we focus our attention on two points of the
function $S(C)$:
\begin{itemize}
\item the point where the system leaves the quasi-equilibrium regime
to enter the aging one, whose coordinates in the plane $(C,T\chi)$ are
$(\qea,1-\qea)$;
\item the point where the system, with a finite $t_w$, converges for
very large times ($t \to \infty$), whose coordinates are
$(0,T\chi_0)$.
\end{itemize}
Note that the first point also represent the equilibrium state into a
single pure state, that can be obtained sending first $t_w \to \infty$
and then $(t-t_w) \to \infty$.  In a model belonging to category A the
two points must have the same height, while we show that $T\chi_0 >
1-\qea$.

We obtain $\qea = 0.37(1)$ from the data of the autocorrelation
function (see Fig.~\ref{corr}) and
\begin{equation}
T \chi_0(t_w) = T \lim_{t-t_w \to \infty} \chi(t,t_w) = 0.75(1) \ ,
\end{equation}
from the data of the response with $t_w=10^6$.  This value should be
considered as a lower bound for $\chi_0$, defined as
\begin{equation}
\chi_0 = \lim_{t_w \to \infty} \chi_0(t_w) \quad ,
\end{equation}
because the data slightly increase with increasing $t_w$~\footnote{If
we take the limits in the reversed order we obtain a relation valid in
a single pure state $T \chi_{\rm eq} = 1-\qea$.}.

\begin{figure}
\begin{center}
\leavevmode
\epsfxsize=0.7 \textwidth
\epsffile{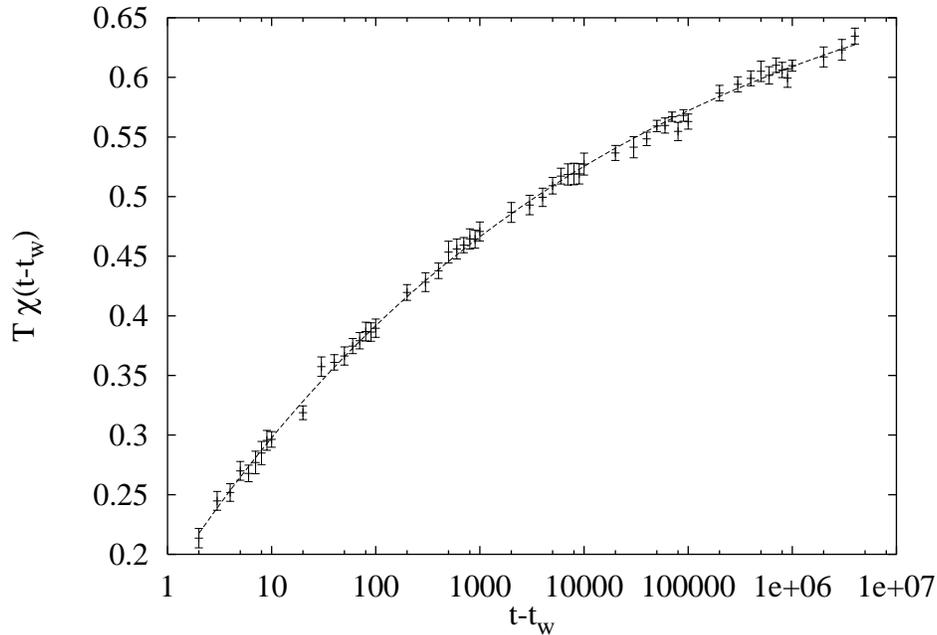}
\end{center}
\caption{Numerical data for the linear response in a system of volume
$32^4$, temperature $T=1.35$ and perturbation $h_0=0.05$ applied after
a waiting time $t_w = 10^6$. The curve is the best fitting power law
(see text).}
\label{chi}
\end{figure}

The extrapolation of the $\chi(t,t_w)$ data is shown in Fig.~\ref{chi}
together with the best fitting power law function:
\begin{equation}
T \chi(t) = 0.75(1) - 0.571(7) t^{-0.102(4)} \ ,
\end{equation}
Even if the exponent is very small, we think our extrapolation to be
very trustworthy because we used more than six time decades and the
fit is very good: $\chi^2/{\rm d.o.f} = \frac{24.2}{54}$, where d.o.f
stands for degrees of freedom.

An exponent so small like the one we found, though in agreement with
previous numerical works~\cite{DYN_H}, could be interpreted as an hint
for a logarithmic law.  We tried to fit the data with some logarithmic
law and we found an asymptotic value greater than the one obtained
with the power law.  So we can assert, with high confidence, than the
value found is a good lower bound for $\chi_0$ and the inequality
holds.

The validity of the inequality $T \chi_0 > 1 - \qea$ confirms that in
the EA model there is a breaking of the replica symmetry.

\section{Conclusions}
\label{S_CONCLUSIONS}

In this work we have shown how the complexity of the frozen phase of a
disordered system can be obtained via the measurements of
autocorrelation and response functions in the out of equilibrium
regime.

We have checked that the generalization of the
fluctua\-tion-dissipation theorem, proposed by Cugliandolo and
Kurchan, is valid for all the models where we have tested it.

In the case of the diluted ferromagnetic model and of the random field
Ising model the off-equilibrium fluctuation-dissipation relation
succeed in predicting the existence of a single pure state at the
equilibrium, even if the out of equilibrium dynamics is very slowened
by the large number of metastable states (like in spin glasses).

We believe that this is a further step in the proof that the method is
robust and we think that this method will give the opportunity of
studying disordered systems with less expensive simulations.

The results obtained with this method and reported here show that the
diluted ferromagnetic and the random field models seem to have no
replica symmetry breaking in the Griffiths phase and especially that
the frozen phase of the Edwards-Anderson model is well described by a
mean-field-like solution.

\section*{Appendix}
\label{S_APPENDIX}

In Sect.~\ref{S_ANA} we obtained the link between the statics and the
dynamics, \ie between the OFDR and the $P(q)$.  With this link we can
now translate to the dynamical framework the usual classification of
complex systems based on the form of replica symmetry
breaking~\cite{MEPAVI}: one step or full replica symmetry breaking.

We will consider always only the positive part of the distribution
function of the overlaps [$P(q) \for q>0$], assuming that the
Hamiltonian is invariant under a global flip of the spins.

\begin{figure}
\begin{center}
\leavevmode
\epsfxsize=\textwidth
\epsffile{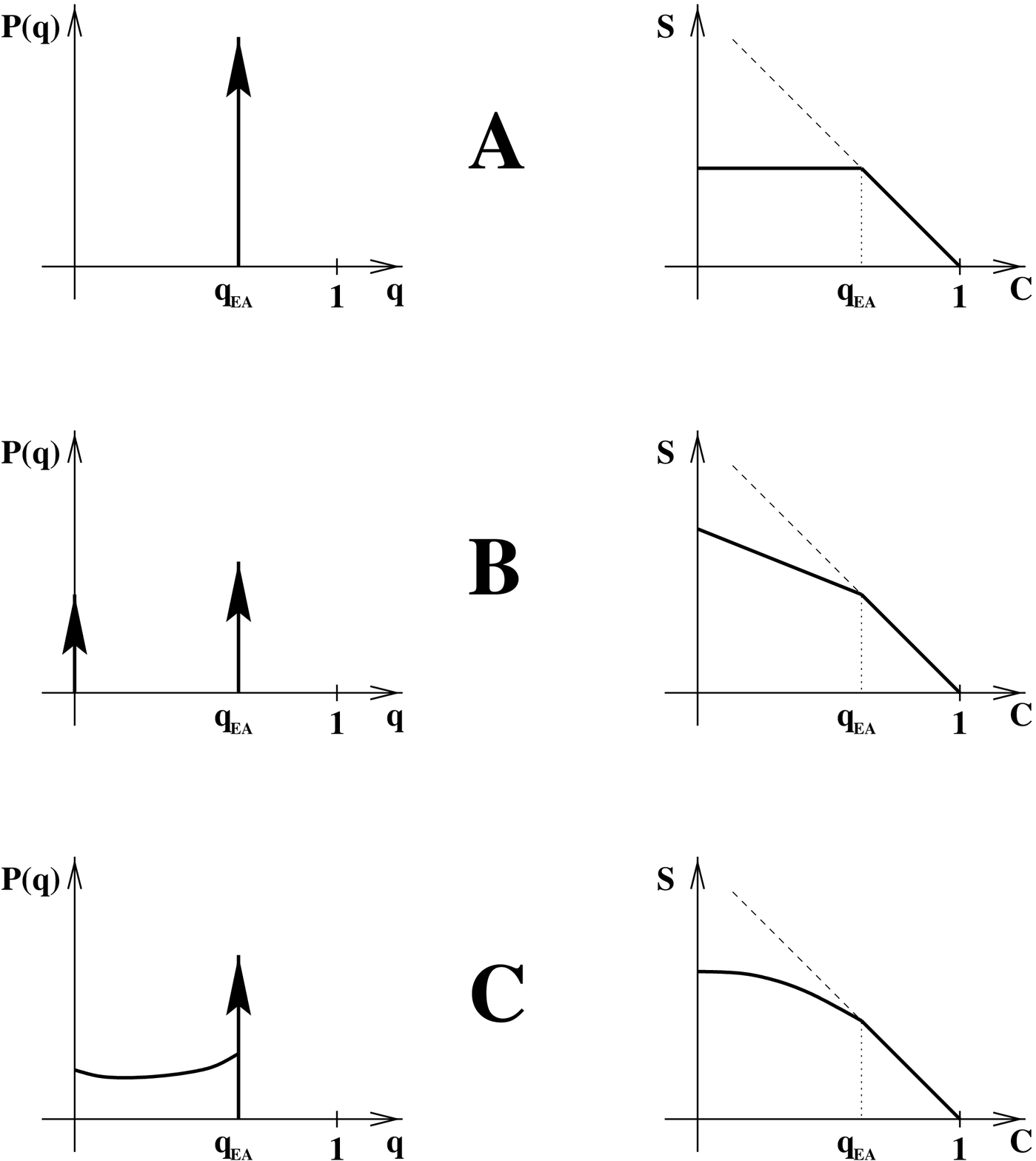}
\end{center}
\caption{A possible model classification based on the function $S(C)$.
The big arrows represent delta functions.}
\label{CLASS}
\end{figure}

In a model whose frozen phase is well described by a single pure state
(no replica symmetry breaking) the distribution function of the
overlap is a single delta function: \eg for the pure ferromagnetic
model with $T<T_c$ we have $P(q)=\delta(q-m_0^2)$, where $m_0(T)$
is the spontaneous magnetization at a given temperature $T$. These
models are the simplest in the sense that their frozen phase can be
described simply giving one parameter.  To such a simple thermodynamic
description correspond an OFDR like the one depicted in
Fig.~\ref{CLASS}A, with an horizontal line in the region
$C<\qea=m_0^2$. We group these models in category A.

The second category (B) groups such models which show a transition
with only one step of replica symmetry breaking ($p$-spin models with
$p>2$ in the mean-field approximation, binary mixtures of soft
spheres~\cite{PARISI_GLASS} and Lennard-Jones
mixtures~\cite{BARRAT_KOB}) and which seem to well describe the real
glass transition. The mean-field version of the $p$-spin model in the
cold phase has a solution with two delta functions, $P(q)=m\ \delta(q)
+ (1-m)\ \delta(q-\qea)$ where $m$ is a function of the temperature
(\eg $m=T/T_c$ in the Random Energy Model), and so their OFDR is given
by two straight lines like the one shown in Fig.~\ref{CLASS}B.

In the category C we include all the models which show an infinite
number of steps of replica symmetry breaking (like the
Sherrington-Kirkpatrick model and the Edwards-Anderson one). They have
a $P(q)$ different from zero in a whole range $q\in [0,\qea]$ with a
delta function on the greater allowed value \qea. These models have an
OFDR like the one shown in Fig.~\ref{CLASS}C.

In any case there is a time region [$(t-t_w)<t_w$ and $C>\qea$], which
we call of {\em quasi-equilibrium}, where the FDT holds and
$S(C)=1-C$.  The differences arise in the {\em aging} regime
[$(t-t_w)>t_w$ and $C<\qea$], where only the models belonging to
category A do not show any aging effect in the response, that is their
integrated response is flat and $S(C)$ is constant.  On the other
hand, models of categories B and C do also respond in the aging
regime, in a way that depends on the value of the waiting time.  This
memory effect is an important feature of a kind of disordered systems
and the method based on the OFDR is able to measure it.

In the aging regime can be naturally defined an effective temperature
via~\cite{CUKUPE}
\begin{equation}
\hat{T} = \frac{T}{X(C)} \qquad \mbox{or} \qquad
\hat{\beta} = X(C) \beta \ .
\end{equation}
This effective temperature is infinite for the models that do not have
memory in the response (cat. A), is finite for the category B models
and takes values in a broad distribution for systems with full replica
symmetry breaking (cat. C).

\end{document}